\documentclass[10pt,twocolumn]{article}
\usepackage[english]{babel}
\usepackage[utf8]{inputenc}
\usepackage[T1]{fontenc}
\usepackage{amsmath}
\usepackage{amsfonts}
\usepackage{graphicx}
\usepackage{esint}
\usepackage{amssymb}
\usepackage{mathtools, nccmath}
\usepackage{lipsum}
\usepackage{gensymb}
\usepackage{physics}
\usepackage{geometry}
\usepackage{float}
\usepackage{soul}
\usepackage{circledsteps}
\usepackage{multirow}
\usepackage{caption} 
\usepackage{hyperref}
\usepackage{times}
\usepackage{multicol}
\usepackage{xcolor}
\usepackage{dblfloatfix}

\hypersetup{
    colorlinks=true,
    linkcolor=blue,     
    urlcolor=blue,
    }
	
\usepackage{color, colortbl}
\definecolor{Gray}{gray}{0.9}

\usepackage{afterpage}
	
\usepackage{fancyhdr}
\fancypagestyle{titlepage}{
\fancyfoot[L]{\textcolor{gray}{* corresponding author}\\\vspace{0.3cm}
}
\fancyfoot[C]{\raisebox{0cm}{\thepage}}}
\usepackage[
    backend=biber,
    style=ieee,
    sorting=none
]{biblatex}

\usepackage{titlesec}
\usepackage{hyperref}
\usepackage{framed}

\begin{document}
\newgeometry{    top=0.4in,
    bottom=1in,
    hmargin=0.5in,}
\begin{titlepage}
\thispagestyle{titlepage}
\end{titlepage}
\begin{figure}[h]
    \centering
    \includegraphics[width=.5\columnwidth]{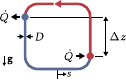}
    \captionsetup{width=1\columnwidth}
    \caption{Schematic of simplified NCL. The low- and high density sections are indicated in red and blue, respectively. The preferential flow direction is indicated by the red arrow.}
    \label{fig:schematic for equation}
\end{figure}

\twocolumn[{%
        \flushright
        \small
        \textbf{The $5^{\text{th}}$ European $\text{sCO}_{2}$ Conference for Energy Systems} \\
        \textbf{March 14-16, 2023, Prague, Czech Republic}\\
        \vspace{0.6cm}
        \LARGE
        \textbf{2023-sCO2.eu.131}\\
        \vspace{1.1cm}
        \centering
        \normalsize
        \textbf{THE STEADY BEHAVIOR OF THE SUPERCRITICAL CARBON DIOXIDE NATURAL CIRCULATION LOOP}
        \vspace{0.5cm}
    \begin{table}[H]
    \centering
           \begin{tabular}{p{1.75cm} c c c}
           &\textbf{Marko Draskic} & \textbf{Benjamin Bugeat} & \textbf{Rene $\text{Pecnik}^{*}$} \\
           &Process $\&$ Energy - 3mE & School of Engineering & Process $\&$ Energy - 3mE\\
           &Delft University of Technology & University of Leicester & Delft University of Technology \\
           & Delft, The Netherlands & Leicester, United Kingdom & Delft, The Netherlands \\
           &\textbf{m.draskic@tudelft.nl} & &\textbf{r.pecnik@tudelft.nl}\\
       \end{tabular}
    \end{table}
    \vspace{0.3cm}
}] 
\paragraph{ABSTRACT}\mbox{}\\
\indent The steady state behavior of thermodynamically supercritical natural circulation loops (NCLs) is investigated in this work. Experimental steady state results with supercritical carbon dioxide are presented for pressures in the range of 80-120 bar, and temperatures in the range of 20-65 $\degree$C. Distinct thermodynamic states are reached by traversing a set of isochors. A generalized equation for the prediction of the steady state is presented, and its performance is assessed using empirical data. Changes of mass flow rate as a result of independent changes of thermodynamic state, heating rate, driving height and viscous losses are shown to be accurately captured by the proposed equation. Furthermore, close agreement between the predicted and measured mass flow rate is found when the measured equipment losses are taken into account for the comparison. Subsequently, the findings are put forward in aid of the development of safe, novel supercritical natural circulation facilities.
\paragraph{INTRODUCTION}\mbox{}\\
\indent When a flow loop is heated at one of its vertical legs and cooled at the other, a natural convection is induced. The flow- and cooling rates of single phase natural circulation facilities are generally orders of magnitude too small to serve a purpose in most industrial applications. However, if the operating fluid is in a thermodynamically supercritical state, considerable flow rates can be obtained due to strong density variations in the vicinity of the critical point. The flow rates generated with these simple systems can be used in settings in which an otherwise moderate flowrate is required, but where problems stemming from leakages, power outages and mechanical noise associated with forced convective flows need to be avoided. For instance, supercritical fluid NCLs can act as reliable, off-grid cooling solutions for nuclear reactors, in case large heat sinks are present. Additionally, these systems can be used for the passive removal of heat from solar heater assemblies, or for the generation of steady, pulseless flows for sensitive experiments. However, as the properties of supercritical media vary greatly with state, the prediction of the steady state of supercritical NCLs for their potential implementation is not straightforward.\\
\indent The steady state of supercritical natural circulation loops has previously been investigated using both numerical and experimental approaches. In the numerical literature, a one-dimensional transient model is most commonly used to predict both the steady and unsteady behavior of the considered loops\hspace{1mm}\cite{ss_predict_sadhu,ss_predict_jain,ss_predict_sharma,ss_predict_pegallapati}, although three-dimensional approaches have also been undertaken\hspace{1mm}\cite{ss_predict_Sarkar}. The mass flow rate of a natural convection loop has been predicted to attain a maximum with varying heating rates\hspace{1mm}\cite{ss_predict_sadhu,ss_predict_jain}. Furthermore, a rise in mass flow rate is expected with increasing loop heights, and increasing channel diameters\hspace{1mm}\cite{ss_predict_sadhu,ss_predict_sharma,ss_predict_Sarkar}. On the contrary, an increase of the loop length is expected to have a limiting effect on the flow rate of the loop\hspace{1mm}\cite{ss_predict_Sarkar} The influence of thermodynamic state on the steady behavior of supercritical NCLs is briefly touched upon in\hspace{1mm}\cite{ss_predict_pegallapati}. Here, increases in both the filling mass and the heating rate are predicted to result in an increase in loop pressure and subsequently loop flow rate for the range of considered parameters. Similar conclusions can be drawn from experimental investigations of NCLs with supercritical media. An increase in mass flow rate with increasing heating rate was first measured by Tokanai et al.,\hspace{1mm}\cite{ss_exp_Tokanai}. The broader range of results presented in Liu et al.\hspace{1mm}\cite{ss_exp_Liu} also show the previously discussed maximum in the mass flow rate with increasing heating. As predicted, an increase in system temperature yields an increase in static pressure at a set charge\hspace{1mm}\cite{ss_exp_Chen,ss_exp_Sadhu_steady}, and an increase in mass flow rate for the considered parameters in the work of Sadhu et al., as shown in\hspace{1mm}\cite{ss_exp_Sadhu_steady}. \\ \\
\indent The above findings only consider and discuss an NCL's sensitivity to changes in specific parameters. A generalized consideration of all variables that affect the steady state is however needed in aid of the reliable design of future facilities. One such correlation of the flow rate of a liquid-like supercritical carbon dioxide NCL, in terms of Grashof and Prantl numbers, was presented by Yoshikawa et al.,\hspace{1mm}\cite{ss_exp_Yoshikawa}. A more elaborate approach was put forward by Swapnalee at al.\hspace{1mm}\cite{theory_Swapnalee}, following the method of Vijayan et al. for single-phase fluids\hspace{1mm}\cite{theory_Vijayan_1994,theory_Vijayan_2002}. In their work, an expression for the mass flow rate is derived from the one-dimensional steady state momentum equation. In order to characterize the distribution of density in the equation \restoregeometry \noindent that follows, the change in loop density has to be expressed as a function of the change in enthalpy in the heater. For this, the relationship between dimensionless density and dimensionless enthalpy introduced by Ambrosini et al.\hspace{1mm}\cite{ambrosini_sharabi} is used. There, the adequate overlap of the dimensionless quantities for a broad range of supercritical pressures makes that a single curve can be used to express the relationship between density and enthalpy. Swapnalee et al.\hspace{1mm}\cite{theory_Swapnalee} use three distinct linear fits of the constitutive curve to express an expected change of density for three separate ranges of subcooling. From this, a straightforward equation for the NCL flow rate follows, which can be expressed in terms of a pipe diameter based Grashof ($Gr_D$) and a Reynolds ($Re_D$) number. A very similar approach is followed in the work of Liu et al.\hspace{1mm}\cite{theory_liu}, where a two-region linear fit of Ambrosini's\cite{ambrosini_sharabi} curve is used to derive an expression for the steady mass flow rate. As the true evolution with state is however continuous, the chosen discrete description of thermodynamic properties is expected to introduce significant errors in the prediction of the flow rate. Additionally, the absence of the characterization of experimental loop minor losses in both works makes that the found relationship between $Gr_D$ and $Re_D$ is configuration specific. As the driving forces are generally limited in comparison with forced convective systems, setup-specific pressure losses in equipment can be expected to considerably reduce the flow rates of NCLs.\\ \\
\indent In this work, a revised generalized equation for the prediction of the steady flow rate of natural circulation loops with supercritical media is proposed. Consequently, the generalized formula is assessed using the experimental results of a supercritical carbon dioxide natural circulation facility at the Process $\&$ Energy laboratory of the Delft University of Technology. In order to find the causes for possible disagreement between theory and experiment, the contributions of state, heating rate, configuration and pressure losses are independently considered.
\paragraph{GENERALIZED FLOW EQUATION}\mbox{}\\ \indent
Away from regions with considerable radial temperature gradients, the flow in a NCL is expected to display behavior similar to that of a developed pipe flow. As such, a generalized equation is sought from the mass- and momentum balance of a one-dimensional flow. Here, a constant-area pipe, and negligible viscous heating are assumed. A geometry that can be described with figure \ref{fig:schematic for equation} is considered. Here, a heater and a cooler are consecutively placed along a closed flow loop. In the figure, the heater and cooler are indicated with red and blue circles, respectively.
 \newpage
\noindent The mass- and momentum balance equations for the considered system are described as a function of streamwise coordinate $s$:
\begin{align}
\frac{1}{A_{\text{cs}}}\frac{\text{d}}{\text{d} s}(\dot{m})&=0, \\
\frac{\text{d}}{\text{d} s}\left(\frac{1}{A_{\text{cs}}^2}\frac{\dot{m}^2}{\rho}+P\right)&=\gamma\rho g-\frac{2f}{D}\frac{1}{A_{\text{cs}}^2}\frac{\dot{m}^2}{\rho}.
\label{eq:momentum}
\end{align}
The coefficient $\gamma(s)\in[-1,1]$ is used to account for the direction of gravity with respect to the flow at coordinate $s$. The rightmost term in equation (\ref{eq:momentum}) describes viscous losses in the system, using dimensionless Fanning factor $f$. Given that the mass flow rate $\dot{m}$ constant at each location, and that $\oint \text{d}(1/\rho)$ and $\oint \text{d}P$ are zero for a closed loop, the path integral of equation (\ref{eq:momentum}) reduces to
\begin{equation}
    \oint \rho g\:\text{d}s=\frac{2}{D A_{\text{cs}}^2}\oint \frac{f\dot{m}^2}{\rho}\:\text{d}s.
    \label{eq:momentum2}
\end{equation}
The equation above is a force balance with contributions solely from the driving buoyancy and viscous losses in the loop. The system is ultimately driven by the density difference $\Delta\rho$ over the vertical section between the cooler and heater with equivalent length $\Delta z$. Here, $\Delta z$ is the vertical distance between the cooler and the heater, if they were to be modeled as point sources and point sinks for heat transfer. In reality, a vertical distribution of $\rho$ is found in both heat exchangers. As such, $\Delta z$ depends on the heat transfer rate, and will attain a value close to the vertical centerline distance. Given the above, equation (\ref{eq:momentum2}) can be rewritten to
\begin{equation}
    \Delta\rho g\Delta zA_{\text{cs}}=\frac{1}{L}\frac{A_{\text{p}}}{A_{\text{cs}}^2}\frac{\dot{m}^2}{2\rho_{\text{m}}}\sum_{i=1}^{n}(f_{\text{i}}L_{\text{i}}),
    \label{eq:momentum3}
\end{equation}
where the viscous loss contributions of all sections $i$ are to be summed. By linearizing the change in density with varying enthalpy at the mean loop temperature $T_{\text{m}}=\int T(s)\text{d}s/L$ and mean loop pressure $P_{\text{m}}$, and assuming constant pressure in all heat transfer equipment, $\Delta\rho$ can be expressed as a function of a change in enthalpy $\Delta h$:
\begin{equation}
    \left.
    \frac{\partial\rho}{\partial h}
    \right\vert_p \Delta hg\Delta zA_{\text{cs}}=\frac{A_{\text{p}}}{A_{\text{cs}}^2}\frac{\dot{m}^2}{2\rho_{\text{m}}}\frac{\Sigma\left(f_{\text{i}}L_{\text{i}}\right)}{L}.
\end{equation}
The use of a single value for $\frac{\partial\rho}{\partial h}\vert_p\Delta h$ to describe $\Delta\rho$ introduces an error as a large thermodynamic space with varying $\frac{\partial\rho}{\partial h}\vert_p$ is traversed. Qualitatively, $\Delta\rho$ is overestimated as $\frac{\partial^2\rho}{\partial h^2}\vert_p$ attains negative values, and $\Delta\rho$ is underestimated when $\frac{\partial^2\rho}{\partial h^2}\vert_p$ is positive. As such, $\Delta\rho$ is overpredicted in the liquid-like region and most notably so in the vicinity of the pseudo-boiling curve, whereas it is underpredicted in the gas-like region. Within the considered thermodynamic range, the magnitude of the error remains within [-10,10] $\%$ for $\Delta h/h_m\leq0.25$. Furthermore, the inherent presence of viscous losses makes that the pressure does not remain constant within heat transfer equipment. However, the pressure losses that are generally obtained at the limited mass flow rates that NCLs can generate have a minimal influence on the local density. Instead, the density varies almost exclusively as a result of changes in enthalpy, induced in the heat exchangers of these systems, and evaluating thermodynamic quantities at constant pressure introduces little error to the prediction.

Using the chain rule, the thermodynamic quantity $\partial\rho/\partial h|_p$ can be rewritten to $\rho_{\text{m}}\beta_{\text{m}}/c_{\text{p,m}}$. Here, all thermodynamic quantities are to be evaluated at $T_{\text{m}}$. Lastly, given that $\Delta h=\dot{Q}/\dot{m}$, $A_{\text{cs}}=\pi D^2/4$ and $A_{\text{p}}=\pi D L$ an equation as a function of design parameters of a supercritical fluid NCL follows:
\begin{equation}
    \dot{m}^3=\frac{\pi^2 g}{32}\cdot{\underbrace{\frac{\rho_{\text{m}}^2\beta_{\text{m}}}{c_{\text{p,m}}}}_{\text{Fluid properties}}}\cdot\underbrace{\dot{Q}\Delta zD^5}_{\text{Configuration}}\cdot\underbrace{\frac{1}{\Sigma(f_{\text{i}}L_{\text{i}})}}_{\text{Viscous losses}}.
    \label{eq:formula}
\end{equation}
Equation (\ref{eq:formula}) expresses expected mass flow rate $\dot{m}$ as a function of a state dependent group of variables, a configuration and geometry specific group, and a viscous loss term. The viscous loss term $\Sigma(f_{\text{i}}L_{\text{i}})$ accounts for both viscous losses in developed sections, and for additional losses in loop equipment and bends. Equation (\ref{eq:formula}) has to be iteratively solved, since the viscous loss term is a function of mass flow rate $\dot{m}$. Furthermore, as the Reynolds numbers for the warm and the cold leg of the system differ at constant $\dot{m}$, their viscous losses have to be solved for independently. The fluid properties of the respective sections can be solved for at $h_{\text{h,c}}=h_{\text{m}}\pm\frac{1}{2}\Delta h|_P$, the value of which follows from the guess for $\dot{m}$. \\ \\
\indent In order to allow for ease of experimental fitting, equation (\ref{eq:formula}) can be rewritten in dimensionless form. For this purpose, dimensionless quantities $Gr_{\text{D}}$ and $Re_{\text{D}}$ are introduced:
\begin{equation}
Gr_{\text{D}}=\frac{\rho_{\text{m}}^2\beta_{\text{m}}}{c_{\text{p,m}}\mu_{\text{m}}^2}\frac{\dot{Q}gD^3}{\dot{m}},\: \: \:Re_{\text{D}}=\frac{\rho_{\text{m}} UD}{\mu_{\text{m}}}.
    \label{eq:GrD, ReD}
\end{equation}
Consequently, the Grashof number can be expressed as a function of the Reynolds number:
\begin{equation}
    Gr_{\text{D}}=2\frac{\Sigma(f_{\text{i}}L_{\text{i}})}{\Delta z}\cdot Re_{\text{D}}^2.
    \label{eq:dimless}
\end{equation}
\noindent In case the pressure losses in loop equipment attain negligible magnitudes, and a fanning factor expression of the form $f=a/Re_{\text{D}}^b$ is used, equation (\ref{eq:dimless}) reduces to
\begin{equation}
    Re_{\text{D}}=\left(\frac{\Delta z}{2aL}\cdot Gr_{\text{D}}\right)^{\frac{1}{2-b}}.
    \label{eq:dimless reynolds}
\end{equation}
Despite having different means to generate a driving force with, both natural and forced convective flows are driven by steady pressure gradients. Hence, ideal fluid friction factor models for forced developed pipe flows are considered for the prediction of $\dot{m}$ in this work. As all values of $Re_{\text{D}}$ for the current empirical data lie within $[10^4,10^5]$, and as the pipe wall surface is hydrodynamically smooth, the simplistic Blasius turbulent friction correlation with constants $a=0.25$ and $b=0.0791$ is used for a comparison with experimental data \cite{zigrang1985review, taler2016determining}. Whilst ideal fluid models perform well when a flow of supercritical carbon dioxide is isothermal \cite{wang2014experimental}, they underpredict the viscous losses when radial temperature gradients are present for the current heat transfer configuration, in which an upward flow is heated and a downward flow is cooled. However, as the loop length far exceeds the total length for which heat is exchanged in the current system, the underprediction of friction by the ideal fluid model is expected to be limited for moderate heating rates. Certainly, a friction model that captures the modulated shear rates of non-isothermal supercritical media is recommended to obtain an accurate prediction of the mass flow rate in smaller, higher power natural circulation loops. 
\begin{figure}[!t]
    \centering
    \includegraphics[width=.97\columnwidth]{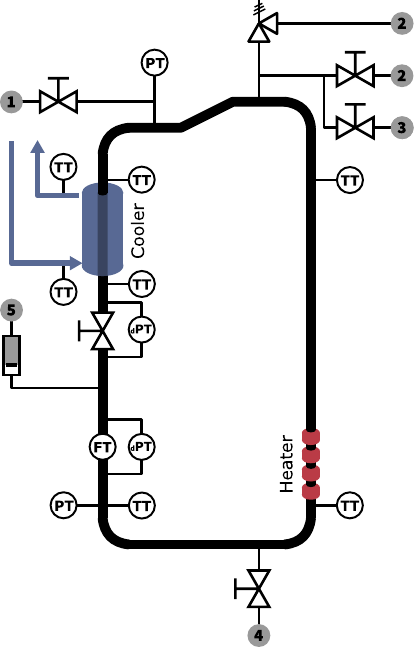}
    \captionsetup{width=1\columnwidth}
    \caption{Schematic depiction of the TU Delft Process $\&$ Energy s$CO_2$ natural circulation loop. As indicated in grey, the system is connected to \Circled{1} a $CO_2$ bottle with dip tube, \Circled{2} a $CO_2$ purge that is connected to the lab's gas vent system, \Circled{3} a vacuum pump, \Circled{4} a drain, and \Circled{5} a nitrogen bottle. The electric heater and annular cooler are indicated in red and blue, respectively.}
    \label{fig:schematic of loop}
\end{figure}
\paragraph{EXPERIMENTAL FACILITY $\&$ METHODOLOGY}\mbox{}\\
\indent The experimental facility designed for- and used in this work is depicted schematically in figure \ref{fig:schematic of loop}. As the heater and cooler are located along the vertical legs of the system, a preferential flow direction prevails. For steady flows, a counter-clockwise circulation is expected in the perspective of the figure. The dimensions of the flow loop and the range of conditions within which it has been designed to operate are specified in table \ref{tab:loop description}. Whereas most of the system is joined using detachable stainless steel tube fittings, EPDM or PTFE is used in components where non-metallic soft seals are required\hspace{1mm}\cite{seals_1}. Heat is supplied to the system using a series of movable electric band heaters. In order to minimize heat losses to the surroundings, the circulation loop is insulated with a 40\hspace{1mm}mm thick annulus of rockwool. The loop is cooled using a tube-in-tube counter-current heat exchanger. Here, the outer annulus is equipped with baffles to aid in the distribution of the coolant. The inlet temperature of the cooler is controlled using a Julabo FP51-SL refrigerated circulator. Whilst mostly simplistic of nature, the loop is also equipped with flow- and state control devices. An adjustable local pressure loss is introduced using a regulating needle valve. Additionally, the volume in the loop can be varied using a 1\hspace{0.5mm}l piston accumulator, indicated below \Circled{5} in the figure. Here, nitrogen is used as the secondary medium.\\

\begin{table}[b]
      \centering
    \begin{tabular}{rlrr}
        \multicolumn{2}{l}{Parameter $\&$ Description} & \multicolumn{1}{r}{Value/Range} & \multicolumn{1}{r}{Unit} \\ \hline \hline
        H  & Loop height & 4.0 & m \\  \hline
        L  & Loop length & 10.0 & m \\  \hline
        D  & Inside diameter & 21.1& mm \\  \hline
        $\Delta z$ & Driving height & $\leq 2.5$& m \\  \hline
        $P$ & Design pressure & $\leq 140$ & bar \\  \hline
       $T$ & Design temperature & $-20\leq T\leq65$ & $\degree$C \\  \hline    
    \end{tabular} 
          \caption{Test loop description}
    \label{tab:loop description}
\end{table}
\begin{figure}[t]
    \centering
    \includegraphics[width=.93\columnwidth]{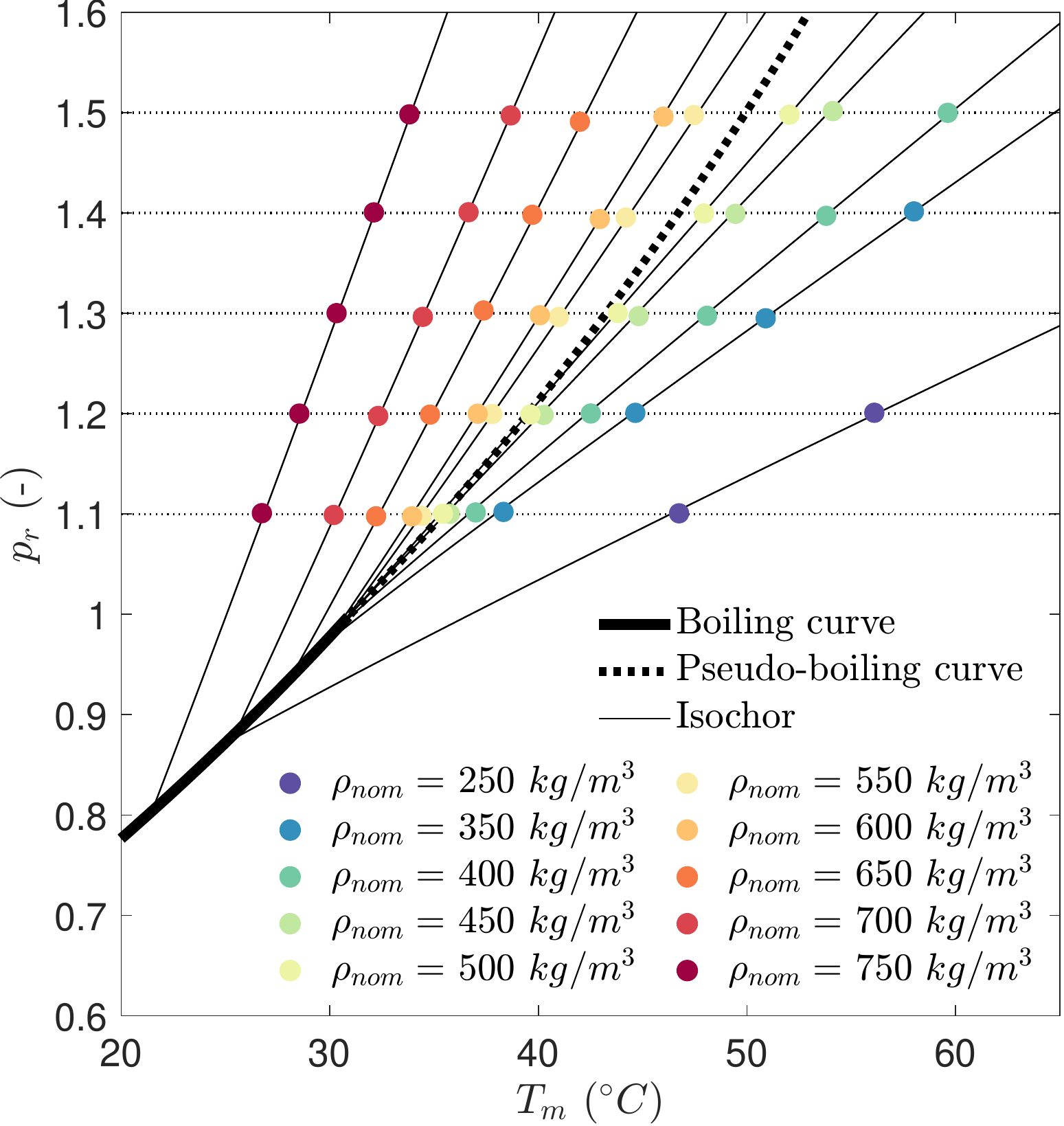}
    \captionsetup{width=1\columnwidth}
    \caption{Lines of constant mass at set system volumes for carbon dioxide at supercritical pressure. All current experimental data points are indicated at the measured $T_{\text{m}}$ and $P_{\text{m}}$ for all considered nominal densities. The boiling- and pseudo-boiling curves are indicated with thick solid and dashed lines, respectively}
    \label{fig:isochors}
\end{figure}
\begin{figure}[t]
        \centering
        \includegraphics[width=.93\columnwidth]{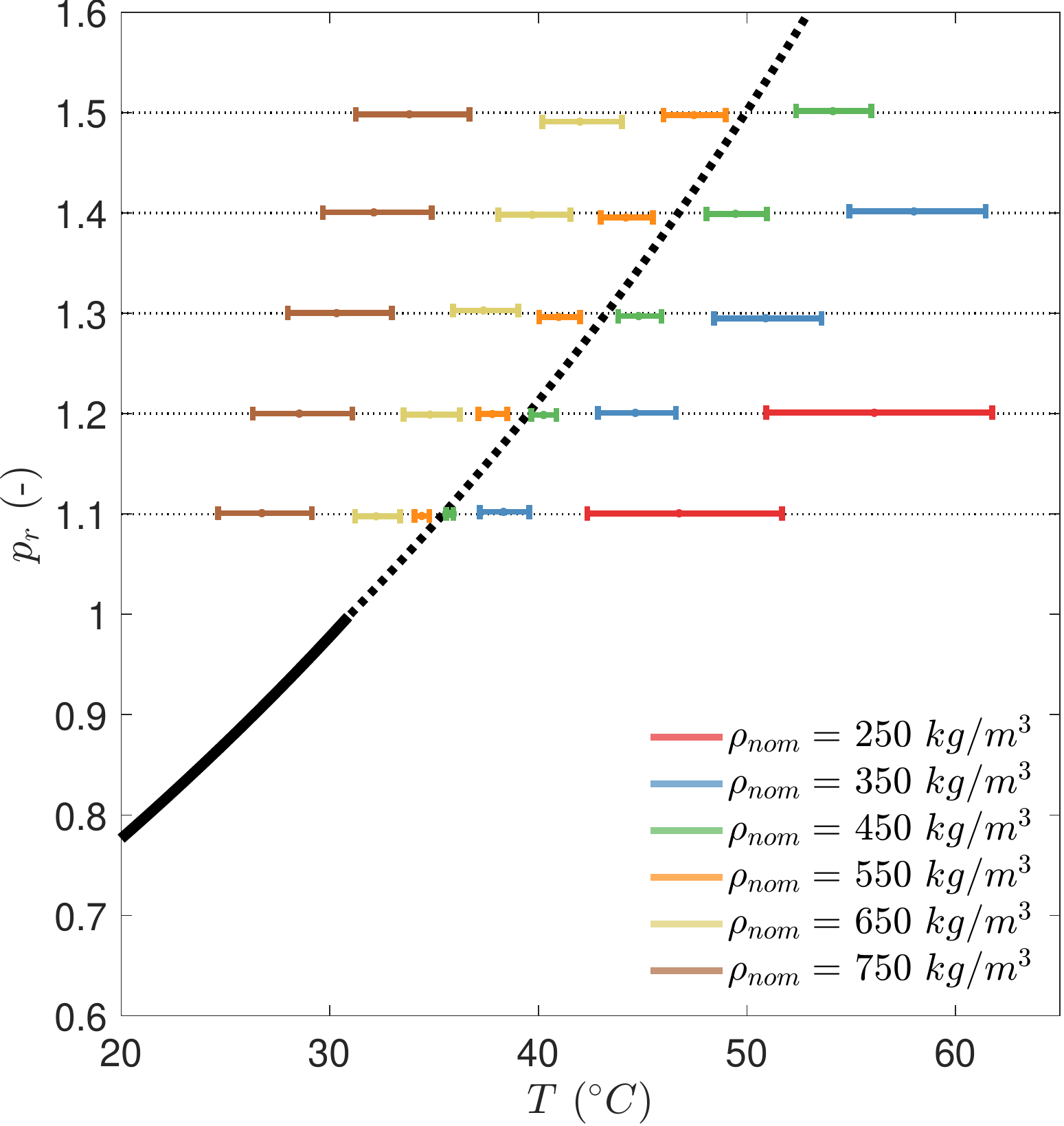}
        \captionsetup{width=1\columnwidth}
        \caption{System temperature ranges for selected loop filling masses, at $\dot{Q}=800$ W, $\Delta z=2.5$ m. The ranges are bound by the greatest and lowest measured temperatures in the loop. The boiling- and pseudo-boiling curves are indicated with solid and dashed lines, respectively.}
        \label{fig:temprange}
\end{figure}
\indent The facility is equipped with a series of transmitters for the continuous monitoring of its performance. Bulk temperatures are measured using PT100 resistance thermometers with a nominal accuracy of $\pm0.1$\hspace{1mm}$\degree C$, which are laterally inserted into the flow. Absolute pressure measurements are taken using welded STS ATM.1st transmitters, with a nominal uncertainty of $\pm0.16$ bar or $0.1\%$. Furthermore, the loop includes a Rheonik RHM08 Coriolis mass flow meter with a nominal uncertainty of $0.2\%$. Finally, two Siemens SITRANS P420 differential pressure sensors were used for the quantification of the viscous losses in both the Coriolis meter and the regulating valve. The transducer data are acquired at up to 10Hz using a NI cRIO-9074. A Labview user-interface for the real-time visualization of the data was developed to complement the data acquisition structure. Here, the interpolation of tabulated thermodynamic properties allows for the live monitoring of various compound quantities.\\

\begin{table}[b]
      \centering
    \begin{tabular}{rlrr}
        \multicolumn{2}{l}{Parameter $\&$ Description} & \multicolumn{1}{r}{Value/Range} & \multicolumn{1}{r}{Unit} \\ \hline \hline
       $\dot{Q}$ & Heating rate & $\leq2$ & kW \\  \hline
       $\rho_{\text{m}}$ & Mass density & $250\leq\rho_{\text{m}}\leq750$ & kg $\text{m}^{-3}$ \\  \hline
       $\text{P}_{\text{m}}$ & Operating pressure & $81\leq \text{P}_{\text{m}}\leq111$ & bar \\  \hline
       $\text{T}_{\text{m}}$ & Operating temperature & $20\leq \text{T}_{\text{m}}\leq 60$ & $\degree$C \\  \hline
    \end{tabular}   
        \caption{Operating range}
    \label{tab:operating range}
\end{table}
\indent In order to fill the loop, the system is first brought to moderate pressures. Consequently, a blow-off valve at the top of the loop is used to purge the system of non-condensable gases. Additionally, a valve at the bottom of the loop is opened to drain the loop of unwanted liquids. After evacuating the system with a vacuum pump, liquid carbon dioxide of a high purity is fed to the system from a cylinder with a dip tube. As the bottle is weighed, the filling mass of the loop is known. Once an equilibrium in pressure is reached between the bottle and the experimental facility, the loop is cooled. As a result, the loop pressure decreases to below the vapor pressure of the bottle, resulting in a flow of carbon dioxide towards the facility. By moderately heating the loop during this cooling step, a natural flow is generated that allows for greater cooling rates, and therewith accelerates the filling process. \\ \\
\indent In this work, the thermodynamic space is explored by traversing a set of isochors. As shown in figure \ref{fig:isochors}, a desired supercritical pressure can be attained for different filling masses at different loop temperatures. During operation, the coolant temperature is adjusted at a constant volume to eventually obtain the appropriate steady state pressure. As such, a constant mean state can be maintained for varying heating rates. The mean state is both continuously and a posteriori evaluated by assuming a linear distribution of $T$ in the heat transfer equipment. The range of states that was therewith reached in this work is listed in table \ref{tab:operating range}, and depicted in figure \ref{fig:isochors}. The distinct thermodynamic states were attained for heating rates of both $400$\hspace{1mm}$W$ and $800$\hspace{1mm}$W$, and are used in discussions of the dependency on thermodynamic state of the mass flow rate and the assessment of the performance of the generalized equation.
  \begin{figure}[t]
    \centering
    \includegraphics[width=1\columnwidth]{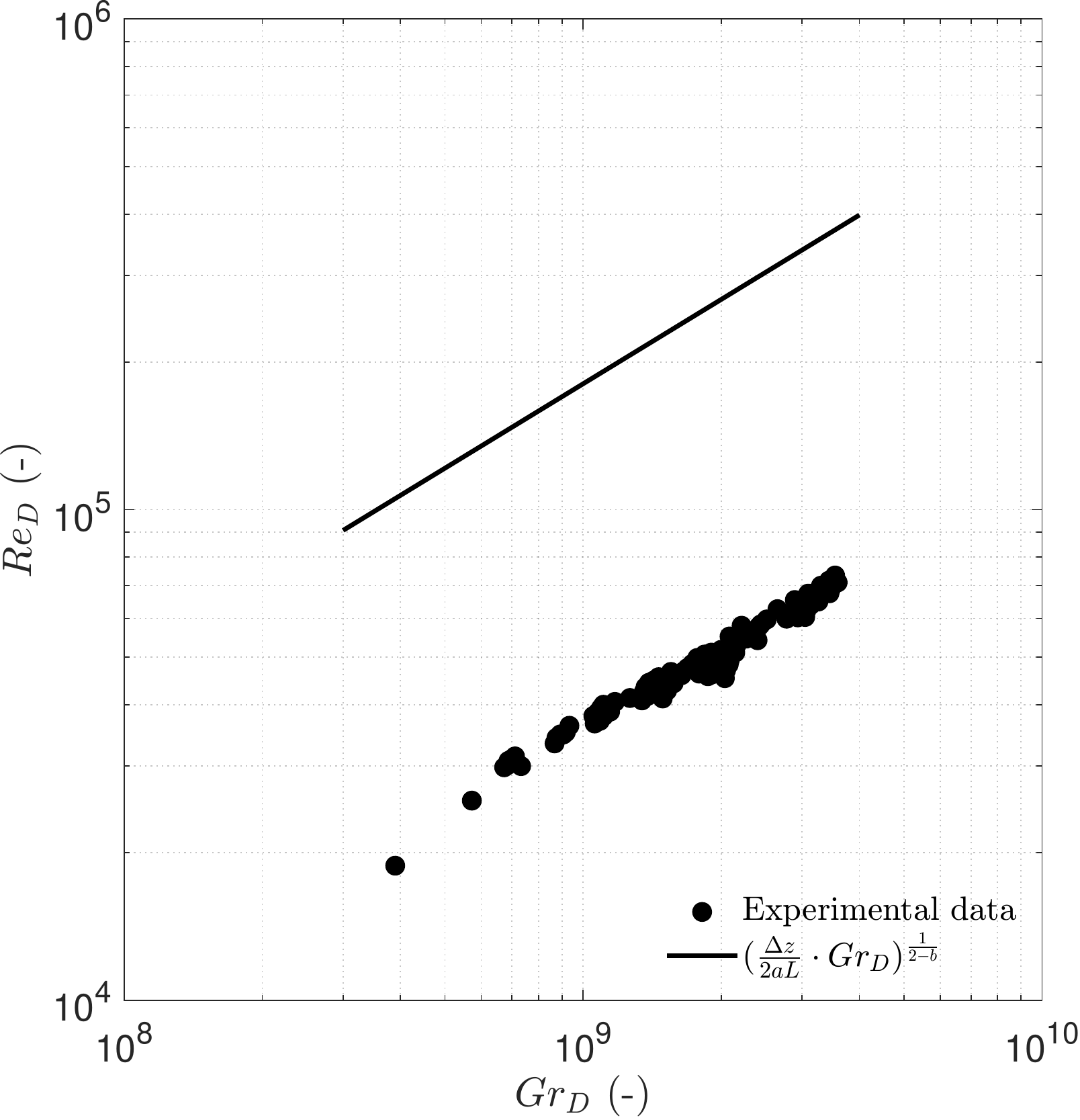}
    \captionsetup{width=\columnwidth}
    \caption{$Re_{\text{D}}$ as function of $Gr_{\text{D}}$ for experimental data in the range of the operating values indicated in table \ref{tab:operating range}. The prediction of equation (\ref{eq:dimless reynolds}) is indicated in the figure with the solid curve.}
    \label{fig:Reynolds Grashof}
\end{figure}
\paragraph{RESULTS $\&$ DISCUSSION}\mbox{}\\
 \indent During the steady operation of the natural circulation loop, the loop temperature distribution varies with thermodynamic state. The loop temperature distribution for an assortment of states within the considered range is depicted in figure \ref{fig:temprange}. The natural flow is driven by moderate temperature gradients, especially in the vicinity of the pseudo-critical line. With increasing pressure beyond the critical point, the pseudo-critical curve gradually transforms from a point of near-discrete phase transition to a gradually increasing region of mild property gradients. Additionally, as the fluid's specific heat near this curve decreases with pressure, less variation of driving temperature gradient is found along isobars of greater magnitudes. Of course, a quantitative assessment of the loop temperature distribution follows from the steady state mass flow rate. If the loop mass flow rate is known, the loop temperature maxima and minima can be obtained using $T_{\text{max,min}}=T_{\text{m}}\pm\dot{Q}/(2\dot{m}c_{\text{p,m}})$. Using the experimental mass flow rate $\dot{m}_{\text{exp}}$, close agreement with experimental data is found for the considered range of thermodynamic states.\\
 
 \indent In this work, equations (\ref{eq:formula}) and (\ref{eq:dimless reynolds}) are proposed for the prediction of steady mass flow rate $\dot{m}$. A comparison of equation (\ref{eq:dimless reynolds}) with experimental data is given in figure \ref{fig:Reynolds Grashof}. The experimental data for this figure is obtained in the absence of the regulating valve depicted in figure \ref{fig:schematic of loop}. As it is for now not yet clear whether the viscous losses in the system have been accurately captured by the used friction model, the predicted mass flow rate is not yet corrected for the measured pressure losses in the mass flow meter. Whereas the measured flow rates are consistently overpredicted in the figure, a comparable trend can be observed between the prediction and the experimental data. In search of generality, the validity of the assumptions and modelling choices made in the derivation of dimensional equation (\ref{eq:formula}) are to be independently tested. As such, the contributions of the individual terms in equation (\ref{eq:formula}) are further investigated in this work.\\

\begin{figure*}[t]
    \centering
    \begin{minipage}[t]{.5\textwidth}
        \centering
        \includegraphics[width=.9\textwidth]{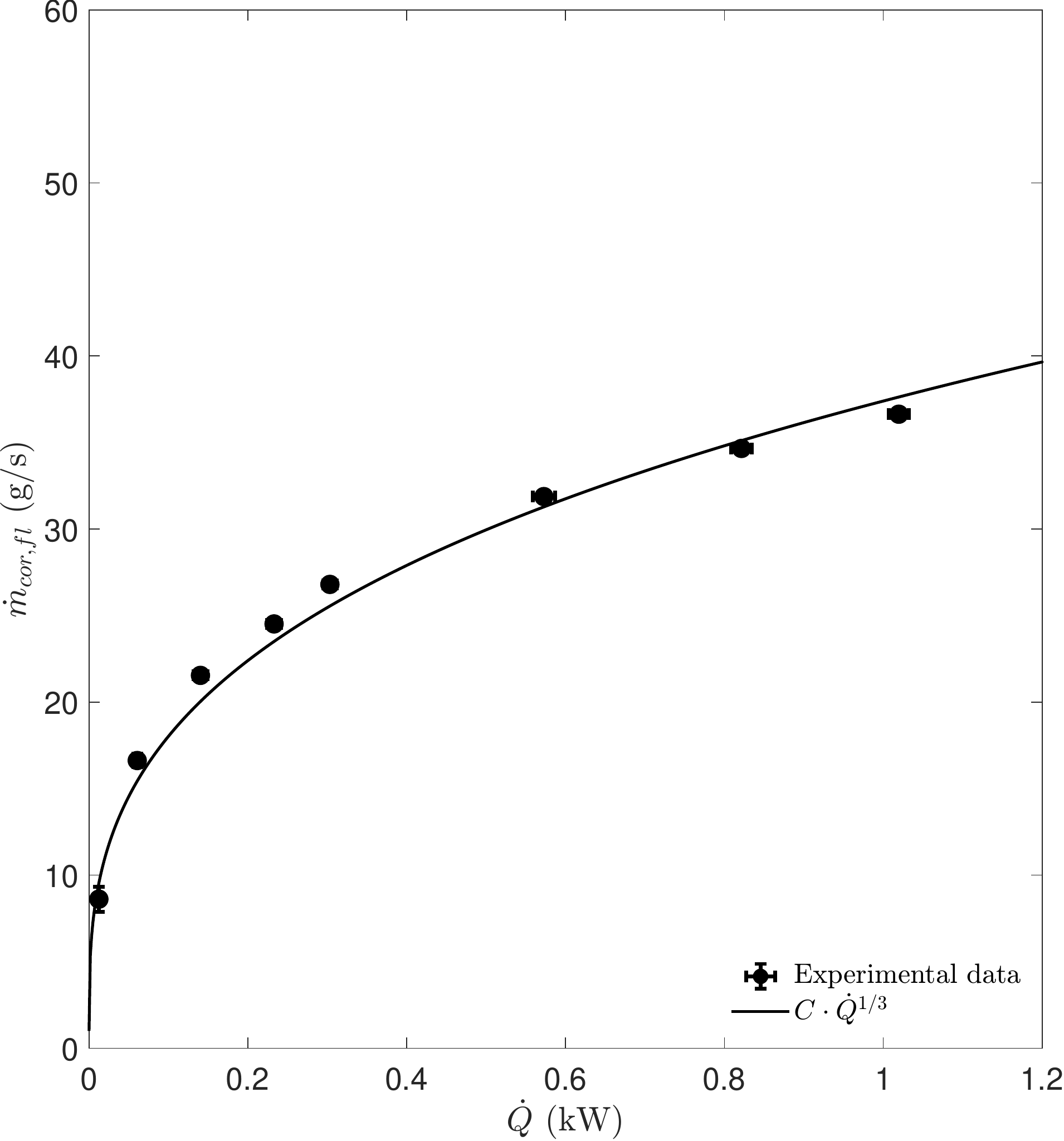}
        \captionsetup{width=0.91\textwidth}
        \caption{Variation of $\dot{m}_{\text{cor,fl}}$ with $\dot{Q}=\dot{m}_{\text{cor,fl}}\Delta h$, at $\rho_{\text{nom}}=700\text{ kg m}^{-3}$, $p_{\text{r}}=1.5$, $\Delta z=2.5 $ m, with 95$\%$ confidence intervals. Constant $C$ is chosen as such that the leftmost data point coincides with the theoretical curve.}
        \label{fig:Heatcomp}
    \end{minipage}%
    \begin{minipage}[t]{0.5\textwidth}
        \centering
        \includegraphics[width=.9\textwidth]{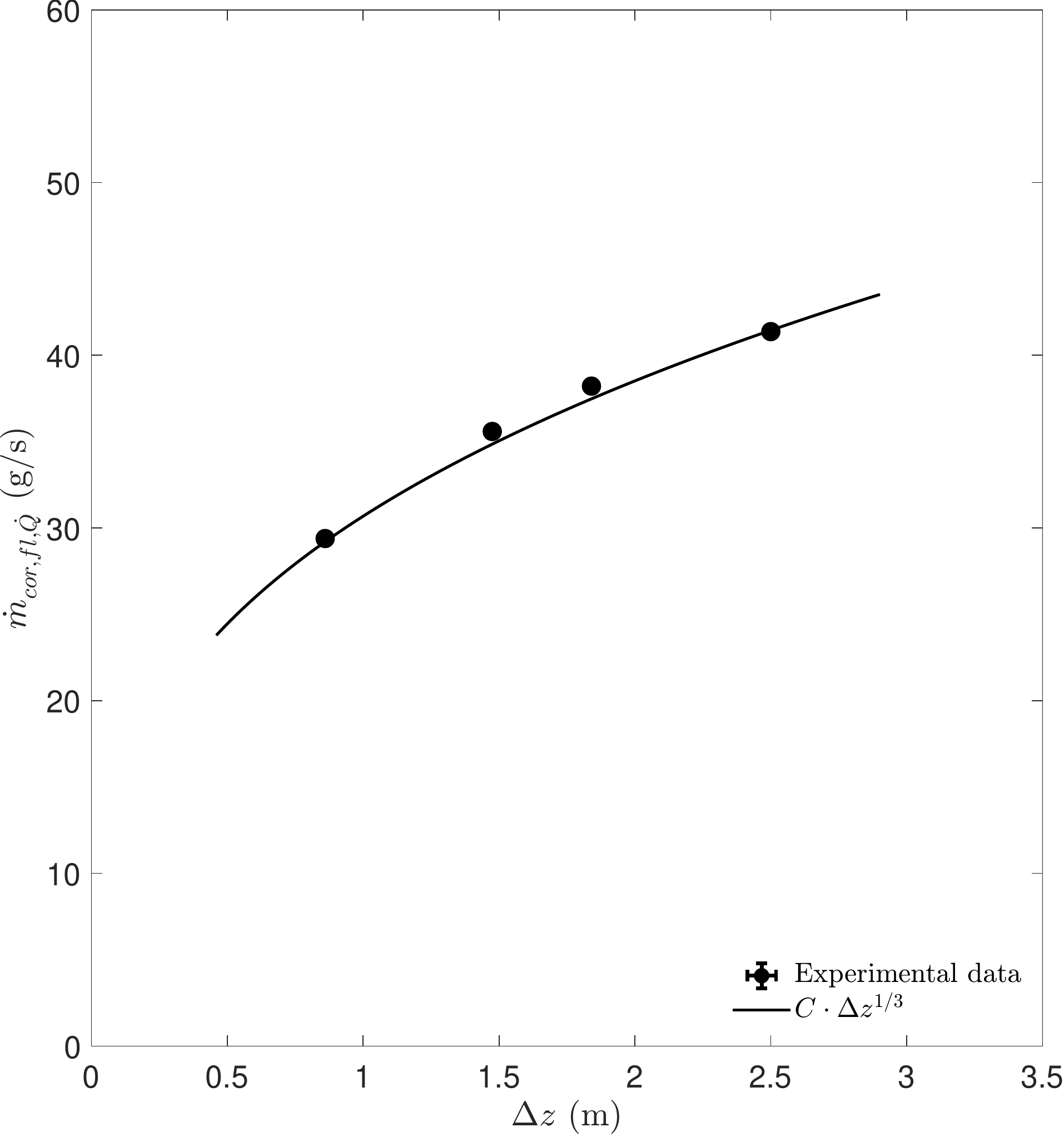}
        \captionsetup{width=0.91\textwidth}
        \caption{Variation of $\dot{m}_{\text{cor,fl,}\dot{Q}}$ with $\Delta z$, at $\rho_{\text{nom}}=700\text{ kg m}^{-3}$, $p_{\text{r}}=1.3$, $\dot{Q}=800$ W, with 95$\%$ confidence intervals. Constant $C$ is chosen as such that the rightmost data point coincides with the theoretical curve.}
        \label{fig:dzcomp}
    \end{minipage}
\end{figure*}

\indent In order to consider the independent contribution of $\dot{Q}$ in equation (\ref{eq:formula}), all other terms have to attain constant values when $\dot{Q}$ is varied. Through variation of the coolant temperature, a constant thermodynamic mean state can be maintained with varying heating rates. As the measured value of $\Sigma(f_{\text{i}}L_{\text{i}})$ is however nonconstant due to variation in $U$, $\dot{m}_{\text{exp}}$ has to be compensated for using
\begin{equation}
\dot{m}_{\text{cor,fl}}=\dot{m}_{\text{exp}}\cdot\left(\frac{C_{\text{fl}}}{\Sigma(f_{\text{i}}L_{\text{i}})_{\text{exp}}}\right)^{1/3}.
\label{eq:comp viscous}
\end{equation}
Here, $\dot{m}$ is assumed to scale with $\Sigma(f_{\text{i}}L_{\text{i}})^{-1/3}$, following equation (\ref{eq:formula}). The value of $C_{\text{fl}}$ should be chosen such that it matches one of the values of $\Sigma(f_{\text{i}}L_{\text{i}})$ within the considered experimental data set. As will be shown later in this work, the experimental uncertainty is the least for $\rho\geq 700\text{ kg m}^{-3}$ and $p_{\text{r}}\geq 1.3$ within the considered range of thermodynamic states. As such, this range of thermodynamic conditions is chosen for the assessment of the individual contributions of $\dot{Q}$, $\Delta z$, and $\Sigma(f_{\text{i}}L_{\text{i}})$. In figure \ref{fig:Heatcomp}, the expected contribution of $\dot{Q}$ is compared to corrected empirical data. Here, the measured increase in heating rate $\dot{Q}=\dot{m}_{\text{cor,fl}}\Delta h$ is used rather than the imposed electrical heating rate $\dot{Q}_{\text{imp}}$, in order to account for heating losses in the system. Close agreement is found between the predicted trend in mass flow rate and the experimental data for the considered range, hence $\dot{m}$ is assumed to scale with $\dot{Q}^{1/3}$ from this point onwards. As such, heating losses can be compensated for in investigations of data sets in which $\dot{Q}$ is to be kept constant using
\begin{equation}
\dot{m}_{\text{cor,}\dot{Q}}=\dot{m}_{\text{exp}}\cdot\left(\frac{\dot{Q}_{\text{imp}}}{\dot{m}\Delta h}\right)^{1/3}.
    \label{eq:correction}
\end{equation}
\begin{figure*}[!t]
    \centering
    \includegraphics[width=.9\textwidth]{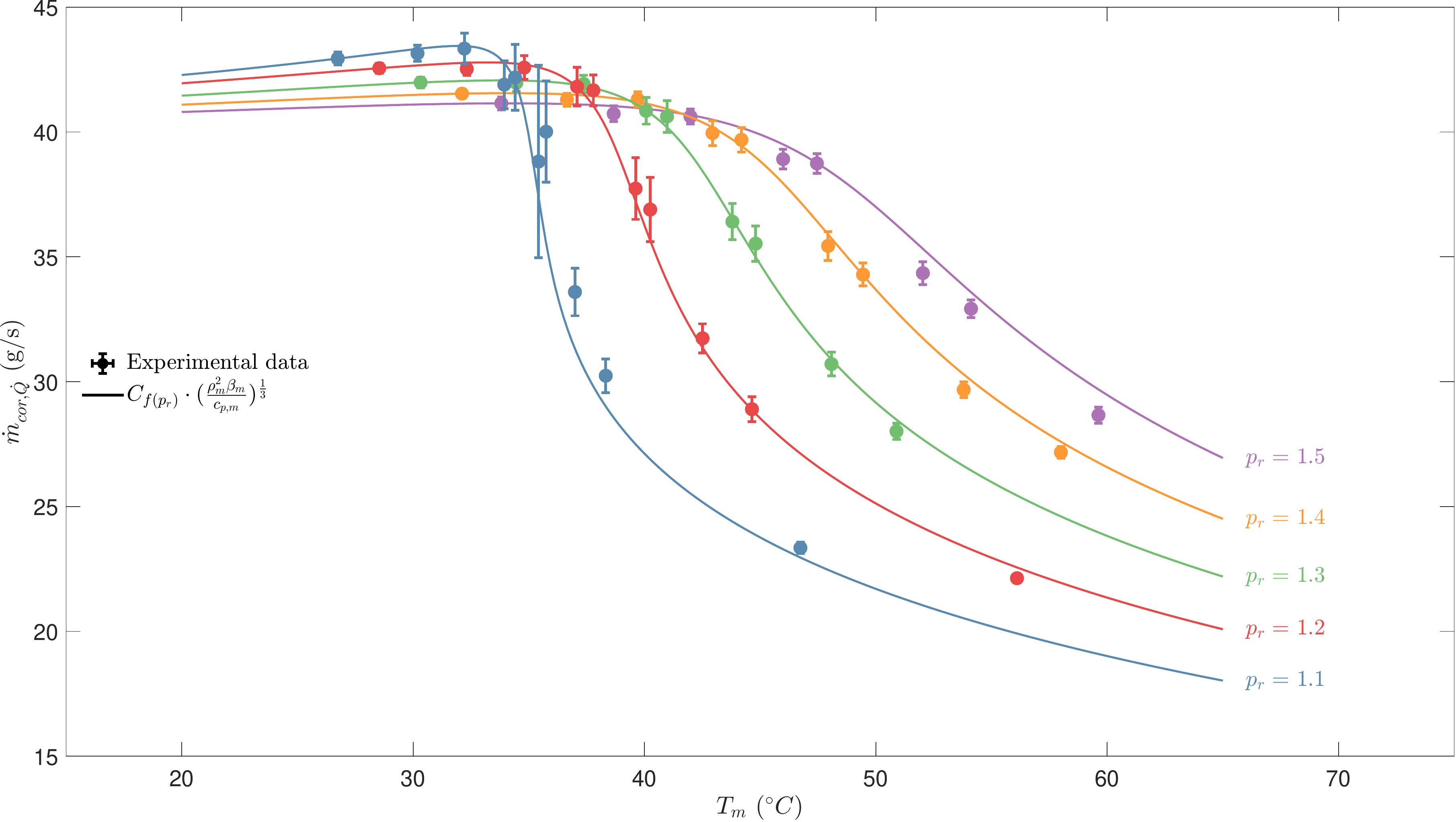}
    \captionsetup{width=0.91\textwidth}
    \caption{Variation of $\dot{m}_{\text{cor,}\dot{Q}}$ with $T_{\text{m}}$ and $P_{\text{m}}$, at $\Delta z=2.5$ m, $\dot{Q}=800$ W, with 95$\%$ confidence intervals. $C_{f(p_{\text{r}})}$ is chosen as such that the leftmost data point for each $p_{\text{r}}$ coincides with the theoretical curve.}
    \label{fig:state contribution}
\end{figure*}
\noindent One such corrected set of empirical data is shown in figure \ref{fig:dzcomp}. The depicted experimental data has been corrected for both variation in viscous losses, and variation in heating losses. Here, the expected change in $\dot{m}$ with variation in $\Delta z$ is compared to data from experiments in which $\Delta z$ is independently varied. Again, close agreement is found between the prediction and experiment.\\ \\
\indent An investigation of the influence of thermodynamic state follows in figure \ref{fig:state contribution}. For the current analysis, the measured value of $\Sigma(f_{\text{i}}L_{\text{i}})$ has been found to vary with pressure, whereas its value remains predominantly constant along each isobar. Hence, the theoretical fluid property contribution of equation (\ref{eq:formula}) is multiplied with $C_{f(p_{\text{r}})}$. The value of this constant is chosen as such that the theoretical curve intersects with the lowest mean temperature data point for each reduced pressure. Heating losses are compensated for using equation (\ref{eq:correction}), and the corrected mass flow rate values $\dot{m}_{\text{cor,}\dot{Q}}$ are shown in the figure. The proposed theoretical contribution of thermodynamic state is found to closely and continuously describe the corrected data for any degree of sub-cooling in the considered range of parameters. Note that the size of the confidence intervals of $\dot{m}_{\text{cor,}\dot{Q}}$ varies greatly with thermodynamic state. The uncertainty in measurements of temperature and pressure is however mostly constant within the current range of experiments. As the sensitivity of enthalpy to temperature however varies with pressure, the uncertainty in enthalpy follows accordingly. This makes that the uncertainty in the determination of the fluid enthalpy used for the correction of heating losses is greatest near maxima of specific heat, hence at the pseudo-critical line at pressures in the vicinity of the critical pressure. Therefore, investigations of individual contributions of equation (\ref{eq:formula}) should be performed away from the pseudo-critical curve, to reduce the uncertainty of the findings. As such, the more liquid-like, high pressure thermodynamic states are considered for these analyses, as previously elaborated on in this work.\\ \\
\indent Finally, the effect of pressure losses caused by equipment is investigated. For this, the joint pressure drop $\Delta p_{\text{e}}$ over the flow meter and the regulating valve section is monitored. Pressure drop $\Delta p_{\text{e}}$ is the summed value of the readings over both differential pressure transmitters indicated with \Circled{\textbf{\scriptsize d\small PT}} in figure \ref{fig:schematic of loop}. A fully developed pipe flow is assumed along the loop, for which the viscous losses are approximated using an ideal fluid model, as elaborated on previously in this work.
\begin{figure}[t]
    \centering
        \centering
        \includegraphics[width=0.97\columnwidth]{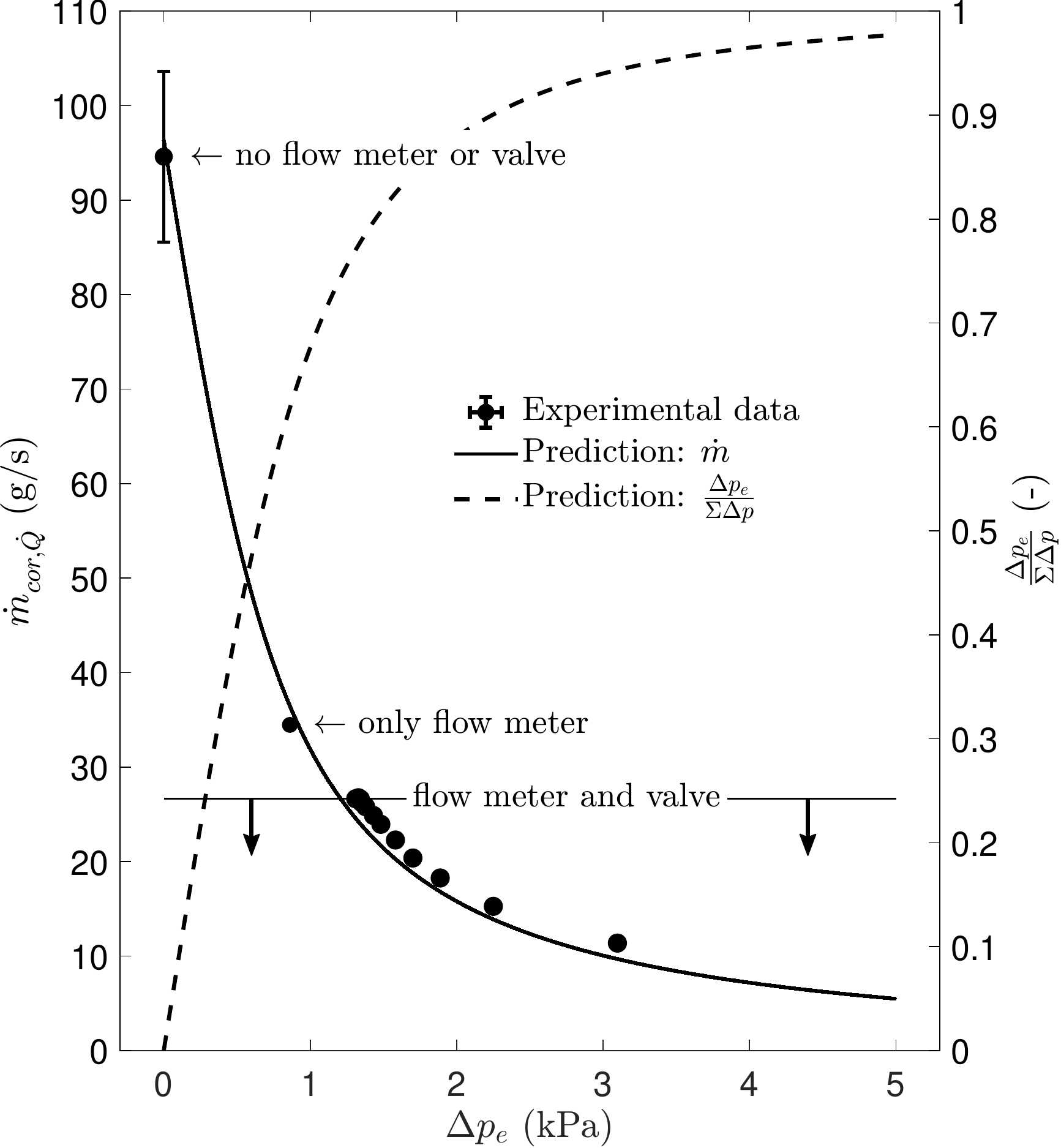}        \captionsetup{width=\columnwidth}
        \caption{Left axis, full line and markers: variation of $\dot{m}_{\text{cor,}\dot{Q}}$ with $\Delta p_{\text{e}}$, at $\rho_{\text{nom}}=730\text{ kg m}^{-3}$, $p_{\text{r}} = 1.1$, $\Delta z=2.5$ m, $\dot{Q}=400$ W, with 95$\%$ confidence intervals. The predictive curve is found by varying the equipment loss term $(fL)_{\text{e}}$ in equation (\ref{eq:formula}), and solving for $\dot{m}$. The leftmost data point has not been corrected to take heating losses into account. Right axis, dotted line: $\Delta p_e$ as a fraction of the estimated total loop pressure losses $\Sigma \Delta p$.}
        \label{fig:friccomp}
\end{figure}
\begin{figure}[t]
        \centering
        \includegraphics[width=\columnwidth]{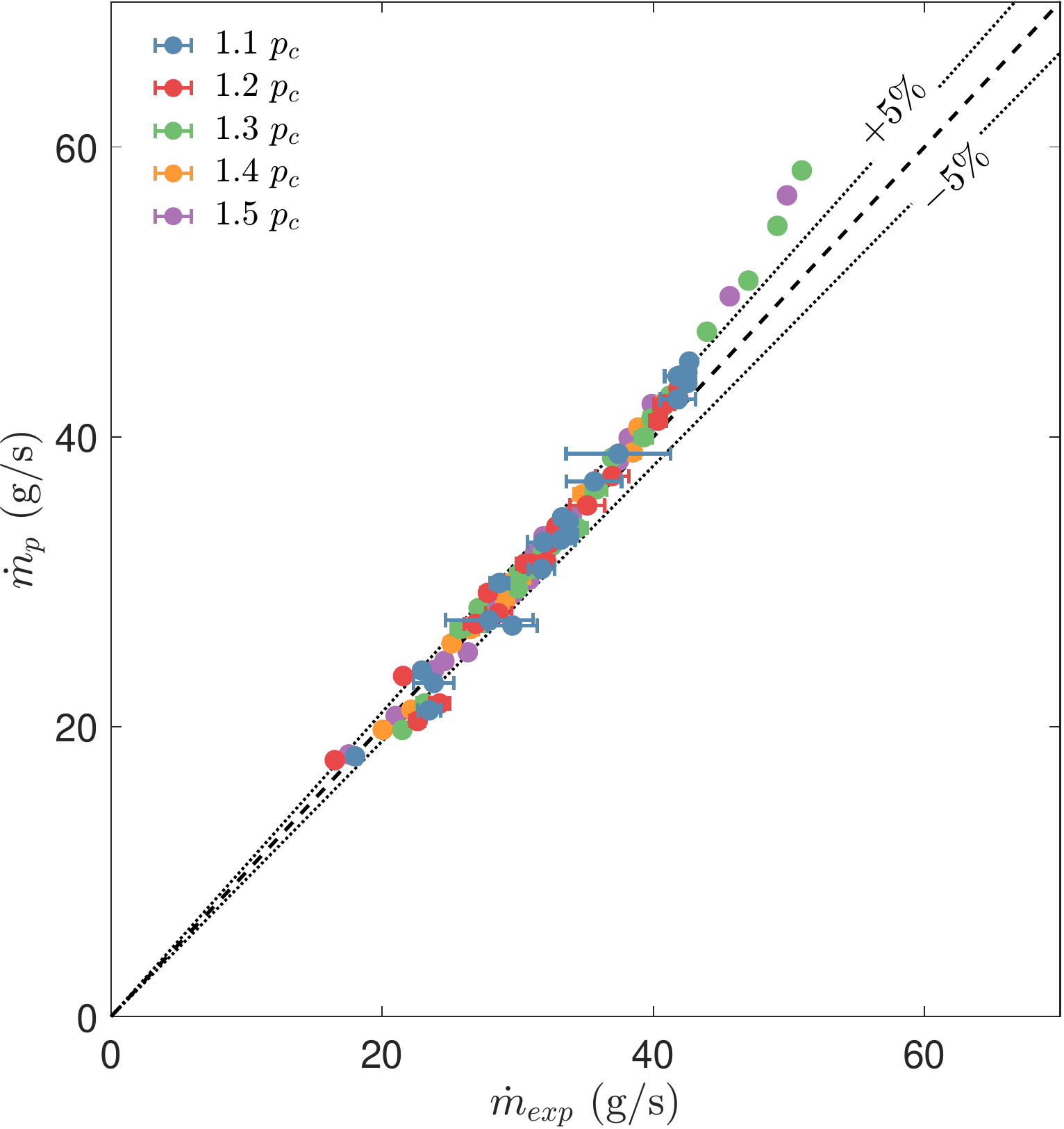}
        \captionsetup{width=\columnwidth}
        \caption{Steady state prediction error with inclusion of measured $\Delta p_{\text{e}}$ into equation (\ref{eq:formula}), at $\Delta z=2.5$ m, with 95$\%$ confidence intervals.}
        \label{fig:error}
\end{figure}
\begin{table*}[!b]
\caption*{\normalsize\textbf{NOMENCLATURE}}
\vspace{-0.4cm}
\begin{framed}
    \footnotesize
    \centering
    \begin{minipage}[t]{.5\textwidth}
    \vspace{-\topskip}
    \vspace{0.4cm}
    \centering
        \begin{tabular}{llr}
        Symbol & Property & Unit\\ \cline{1-3}
        \rowcolor{Gray}
        \multicolumn{3}{c}{\textit{Symbols}} \\
        $a$ & Constant in eq (\ref{eq:dimless reynolds}) & - \\
        $A_{\text{cs}}$ & Cross sectional area $(\pi D^2/4)$& $\text{m}^{2}$ \\
        $A_{\text{p}}$ & Pipe wall area $(\pi DL)$& $\text{m}^{2}$\\
        $b$ & Constant in eq (\ref{eq:dimless reynolds}) & - \\
        $c_{\text{p}}$ & Specific heat & J $\text{kg}^{-1}\text{K}^{-1}$ \\
        $C_{\text{fl}}$ & Constant in eq. (\ref{eq:comp viscous}) & - \\
        $D$ & Internal pipe diameter & m \\
        $f$ & Fanning friction factor & - \\
        $F$ & Force & N \\
        $g$ & Gravitational acceleration & $\text{m s}^{-2}$ \\
        $Gr_{\text{D}}$ & Grashof number, eq (\ref{eq:GrD, ReD}) & - \\
        $h$ & Specific enthalpy &  J $\text{kg}^{-1}$\\
        $H$ & Total loop height & m \\
        $L$, $L_{\text{i}}$ & Loop length, section length & m \\
        $\dot{m}$ & Mass flow rate & kg $\text{s}^{-1}$ \\
        $P$ & Pressure (absolute) & Pa \\
        $\dot{Q}$ & Heating rate & W \\
        $s$ & Streamwise coordinate & m \\
        $T$ & Temperature & K \\
        $U$ & Streamwise velocity component & m $\text{s}^{-1}$ \\
        $\Delta z$ & Vertical point-source heater-cooler distance & m \\
        \end{tabular}
    \end{minipage}%
    \begin{minipage}[t]{0.5\linewidth}
    \vspace{-\topskip}
    \vspace{0.4cm}
    \centering
        \begin{tabular}{llr}
            Symbol & Property & Unit\\ \cline{1-3}
        \rowcolor{Gray}
        \multicolumn{3}{c}{\textit{Greek symbols}} \\
        $\beta$ & Volumetric expansivity & $\text{K}^{-1}$ \\
        $\gamma$ & Directional coefficient in eq (\ref{eq:momentum}) & - \\
        $\mu$ & Dynamic viscosity & Pa s\\
        $\rho$ & Density & kg $\text{m}^{-3}$ \\
        \rowcolor{Gray}
        \multicolumn{3}{c}{\textit{Subscripts}} \\
        c & \multicolumn{2}{l}{Value taken at temperature and pressure of cold leg} \\
        $\text{cor,}\dot{Q}$ & Corrected for variation in $\dot{Q}$ \\
        cor,fl & Corrected for variation in $\Sigma(f_{\text{i}}L_{\text{i}})$ \\
        crit & \multicolumn{2}{l}{Value at critical point} \\
        D & \multicolumn{2}{l}{Pipe inner diameter as characteristic lengthscale} \\
        e & \multicolumn{2}{l}{Value at specified equipment} \\
        exp & \multicolumn{2}{l}{Experimental value} \\
        h & \multicolumn{2}{l}{Value taken at temperature and pressure of hot leg} \\
        imp & \multicolumn{2}{l}{Imposed value} \\
        m & \multicolumn{2}{l}{Value taken at loop mean temperature and pressure} \\
        nom & \multicolumn{2}{l}{Nominal/intended value} \\
        p & \multicolumn{2}{l}{Predicted value} \\
        r & \multicolumn{2}{l}{Reduced value, with respect to value at critical point} \\
        \end{tabular}
    \end{minipage}
\end{framed}
\end{table*}
Figure \ref{fig:friccomp} shows the loop mass flow rate as a function of the experimental values of $\Delta p_{\text{e}}$, and the relative magnitude of $\Delta p_{\text{e}}$ with respect to the estimated total viscous pressure losses in the system. The regulating valve is present in all but two data points of the current analysis, for which it is removed from the supercritical NCL. In the leftmost data point in the figure, both the valve and the mass flow meter are removed from the experimental facility. Here, the mass flow rate is estimated from the imposed heating rate and the measured enthalpy increase over the heater, i.e. $\dot{m}_{\text{exp}}=\dot{Q}_{\text{imp}}/(h_{\text{h,exp}}-h_{\text{c,exp}})$. The high sensitivity to temperature in deducing enthalpy makes that the uncertainty for this data point is significantly larger than for the data for which a mass flow meter is still present. Furthermore, as heating losses cannot be reliably estimated for this data point, they are not compensated for. The predictive curve is obtained by adding a fictitious equipment loss term $(fL)_{\text{e}}=(\Delta p_{\text{e}}D)/(2\rho_{\text{c}} U_{\text{c}}^2)$ in the viscous loss term in equation (\ref{eq:formula}). For each value of $\Delta p_{\text{e}}$, the distribution of the viscous losses in the system has to be iteratively solved for. As such, the presented theoretical curve in figure \ref{fig:friccomp} is state- and configuration dependent and therewith only applicable to the current analysis.\\

\indent The theoretical curve is found to closely resemble the experimental data, indicating that the used viscous model accurately captures the viscous losses for the current configuration and heating rate. The value of the predictive curve at a zero value of $\Delta p_{\text{e}}$ corresponds to the expected mass flow rate in case no equipment losses are assumed. A comparison of the two leftmost points in the figure shows that a threefold decrease in mass flow rate is the direct consequence of the inclusion of a flow meter in the experimental loop. The difference in mass flow rate is of similar magnitude as the shift in figure \ref{fig:Reynolds Grashof}, in which equipment pressure losses were not taken into account. As can be seen from figure \ref{fig:friccomp}, the losses in the flow meter alone are estimated to exceed the regular viscous losses in the loop. For all the considered experimental data points of this work, the total loop viscous losses are dominated by the equipment losses of a single Coriolis transmitter. The associated loss of flow rate is expected to be even more prevalent in facilities with less simple geometries, in which the flow is forced through a greater amount of instruments, or past series of turbine blades. Hence, the a priori characterization of the equipment minor losses is highly recommended for an accurate prediction of the steady mass flow rate of yet to be developed supercritical NCLs, which are generally inflexible to changes in maximum heat throughput. Without the proper portrayal of such losses, equation (\ref{eq:formula}) will only serve as a qualitative measure of the sensitivity of $\dot{m}$ to changes in thermodynamic state, $\Delta z$ and $\dot{Q}$.\\

\indent Figure \ref{fig:error} shows the error in the prediction of the experimental steady state mass flow rate $\dot{m}_{\text{exp}}$, if the measured $\Delta p_{\text{e}}$ is included in the viscous loss term of equation (\ref{eq:formula}). As shown in the figure, close agreement is found with the predicted mass flow rate $\dot{m}_{\text{p}}$ for the majority of the data.\\

\indent For larger mass flow rates, the experimental flow rate is increasingly overpredicted. A likely cause for the above is an underprediction of the viscous losses in the flow in the non-adiabatic sections of the setup by the ideal fluid friction model used in the comparison. As qualitatively described in the work of Wahl et al.\hspace{1mm}\cite{wahl_buoyancy}, the alignment of the direction of both forced- and natural convection can result in near-wall velocities that are greater than in an adiabatic setting. For flows of supercritical media, the above applies to downward cooled and upward heated pipe flows. Since such alignment is present in both heat transfer configurations of the investigated loop for the preferential flow direction, enhanced buoyancy-aided shear is expected in- and directly downstream of the cooler and heater. The deviation from ideal behavior is expected to increase for thermodynamic states that are more liquid-like, and at greater heating rates \cite{fric_correlations_fang}, at which larger mass flow rates are also expected. As such, the increasingly large overprediction shown in figure \ref{fig:error} is expected. Closer agreement between equation \ref{eq:formula} and empirical data would be found if a more complex friction model that captures the non-ideal modulation of shear in the heat exchangers would be used for the comparison. However, if aware of its caveats, a simple ideal fluid friction model can already be used to yield fair and quick predictions of the mass flow rate for a large parameter space, as can be deducted from figure \ref{fig:error}.
\paragraph{CONCLUSIONS}\mbox{}\\
\indent The steady state behavior of a natural circulation loop that employs thermodynamically supercritical carbon dioxide was experimentally investigated in this work. The experiments were conducted using a novel facility at the Process $\&$ Energy laboratory of the Delft University of Technology. Distinct empirical data points were obtained by varying the system's filling mass, its heating rate, and the temperature of the coolant. The experimental data was compared to a newly proposed generalized equation for the prediction of the steady state mass flow rate of supercritical NCLs. Close agreement between the predicted flow rate and empirical data is found within the considered range of operating conditions and parameters, if equipment pressure losses were accounted for in the generalized equation. Furthermore, the sensitivity to independent changes in heating rate, differential heating height, viscous losses, and thermodynamic state was shown to be accurately captured. Any limitations of the current equation and the assumptions made in its derivation were furthermore discussed, to aid in the safe and reliable development of future large-scale supercritical NCL systems.
\newpage
\paragraph{ACKNOWLEDGEMENTS}\mbox{}\\
\indent This work was funded by the European Research Council grant no. ERC-2019-CoG-864660, Critical.
\printbibliography[title=\normalsize{REFERENCES}]
\end{document}